\documentclass[conference]{IEEEtran}

\normalsize

\ifCLASSINFOpdf

\else

\fi

\usepackage{url,color,cite}
\usepackage{caption}
\usepackage[pdftex]{graphicx}
\usepackage[cmex10]{amsmath}
\usepackage{amssymb}
\usepackage{amsthm}
\usepackage{float}
\usepackage{epstopdf}
\usepackage{hyperref}
\usepackage{algorithm,algpseudocode}
\usepackage{algpseudocode}
\usepackage{pifont}
\usepackage{subcaption}
\usepackage{cleveref}
\usepackage{multirow}

\newtheorem{theorem}{Theorem}
\theoremstyle{definition}

\theoremstyle{definition}

\theoremstyle{definition}
\newtheorem{corollary}{Corollary}
\IEEEoverridecommandlockouts
\date{}
\title{A Converse Bound for Cache-Aided Interference Networks }

\author{\thanks{This work was supported in part by the European Union’s Horizon $2020$ research and innovation programme under the Marie Skłodowska-Curie grant agreement No $690893$ and a grant from the Egyptian National Telecommunications Regulatory Authority (NTRA).}
\IEEEauthorblockN{Antonious M. Girgis\IEEEauthorrefmark{1}, Ozgur Ercetin\IEEEauthorrefmark{2}, Mohammed Nafie\IEEEauthorrefmark{1}\IEEEauthorrefmark{3}, and Tamer ElBatt\IEEEauthorrefmark{4}\IEEEauthorrefmark{3}
        }
    \IEEEauthorblockA{\IEEEauthorrefmark{1}Wireless Intelligent Networks Center (WINC), Nile University, Cairo, Egypt\\
\IEEEauthorrefmark{2} Faculty of Engineering and Natural Sciences, Sabanci University, Turkey.\\
        \IEEEauthorrefmark{3} Electronics and Communications Engineering Dept., 
Faculty of Engineering, Cairo University, Egypt.\\
        \IEEEauthorrefmark{4} Computer Science and Engineering Dept., The American 
University in Cairo, Egypt. \vspace{-6mm}
    }}

\begin{document}
\maketitle

\begin{abstract}
In this paper, an interference network with arbitrary number of transmitters and receivers is studied, where each transmitter is equipped with a finite size cache. We obtain an information-theoretic lower bound on both the peak normalized delivery time (NDT), and the \textit{expected} NDT of cache-aided interference networks with uniform content popularity. For the peak NDT, we show that our lower bound is strictly tighter than the bound in the literature for small cache sizes. Moreover, we show that the feasibility region on the expected NDT is bigger than that of the peak NDT.     
\end{abstract}


\section{Introduction}
The exponential growth of on-demand video streaming causes an inevitable burden on wireless networks during the peak hours. Caching is a promising solution to alleviate this problem by pushing the popular data content into cache memories at edge nodes during the off-peak hours, where the network resources are under-utilized. Hence, in the peak hours when the network is congested, caches can be exploited to serve the receivers requests with a significant improvement in the system performance. The rule of caching in interference networks is studied in~\cite{maddah2015cache,kakar2017fundamental,sengupta2016cache,tandon2016cloud,zhang2018fundamental,naderializadeh2017fundamental,hachem2016degrees,xu2017fundamental,girgis2017decentralized,girgis2018NDT}. In~\cite{maddah2015cache}, the degrees of freedom (DoF) of a $3\times 3$ interference network  with cache-equipped transmitters was studied. In~\cite{tandon2016cloud}, the normalized delivery time (NDT) which defines the delivery latency is introduced as a performance metric to study the fog radio access network (F-RAN) with two transmitters and two receivers. The work in~\cite{naderializadeh2017fundamental,hachem2016degrees,xu2017fundamental,girgis2017decentralized,girgis2018NDT} studied interference networks and F-RANs with caches at both transmitters and receivers.

\subsection{Contribution}
In this work, we study a cache-aided interference network with arbitrary number of transmitters and receivers. In contrast to the prior works, we study the information theoretic limits of both the peak NDT and the \textit{expected} NDT under uniform content popularity, where we derive a lower bound on both the expected NDT and the peak NDT for uncoded placement. To the best of the authors knowledge, this paper is the first work discussing the \textit{expected} NDT, since all previous works only study the peak NDT for the worst-case demand. Perhaps, the closest to our work is~\cite{sengupta2016cache}, where the authors derive a lower bound on the peak NDT for uncoded placement schemes. In this paper, we provide a tighter bound for the small cache sizes, i.e., when the cache at each transmitter can store at most the half of the library. Moreover, we show that the feasibility region of the expected NDT is bigger than the feasibility region of the peak NDT. Hence, the achievable schemes designed for the peak NDT should be improved to work with the general demands in which receiver demands are not distinct.

\begin{figure}[t]
\centering{\includegraphics[scale=0.22,trim=6 6 6 6,clip]{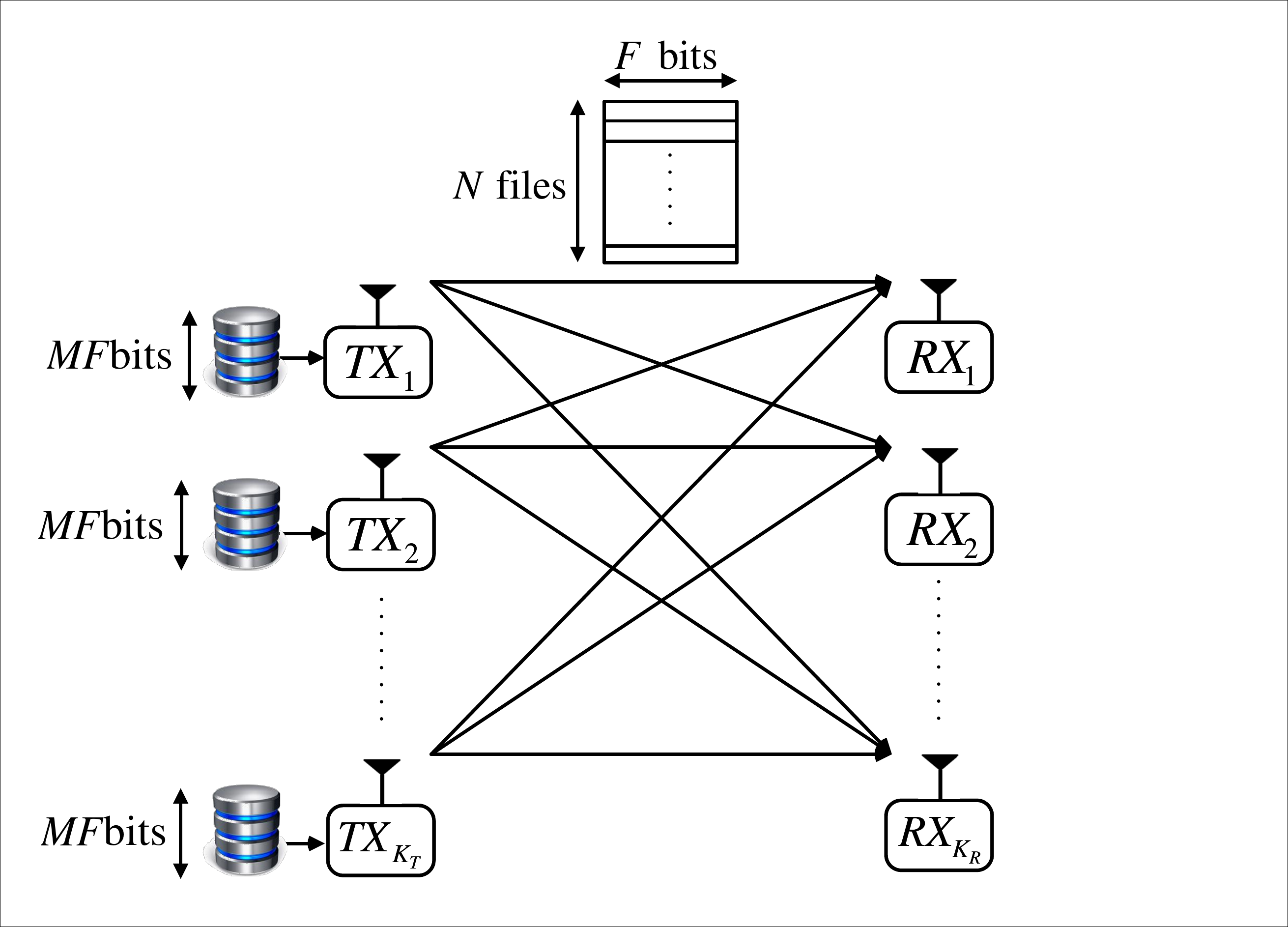}}
\caption{Cache-aided interference network with $K_T$ transmitters and $K_R$ receivers.}
\label{fig1}
\vspace{-5mm}  
\end{figure}  

\section{System Model}

We consider an interference network of $K_T$ transmitters connected to $K_R$ receivers over a Gaussian channel as depicted in Figure~\ref{fig1}. There is a content library of $N$ files, $\mathcal{W}\triangleq\left\{W_1,\cdots,W_N\right\}$, each of size $F$ bits. Each receiver can randomly and independently request a file from the library according to uniform distribution $\lbrace p_i=\frac{1}{N}\rbrace$ for $i\in\left[N\right]$. Each transmitter $\text{TX}_i$, $i\in\left[K_T\right]$, has a local cache memory $Z_i$ of size $MF$ bits, where $\mu=M/N$ refers to normalized cache size. The system operates in two separate phases, a \textit{placement phase} and a \textit{delivery phase}. In the placement phase, the transmitters have access to the content library $\mathcal{W}$, and hence, each transmitter fills its cache memory as an arbitrary function of the content library $\mathcal{W}$ under its cache size constraint. We maintain that the caching functions are designed without any prior knowledge of the future receivers demands and the channel coefficients between transmitters and receivers.

\begin{figure}[t!]
\centering{\includegraphics[scale=0.24,trim=4 4 4 4,clip]{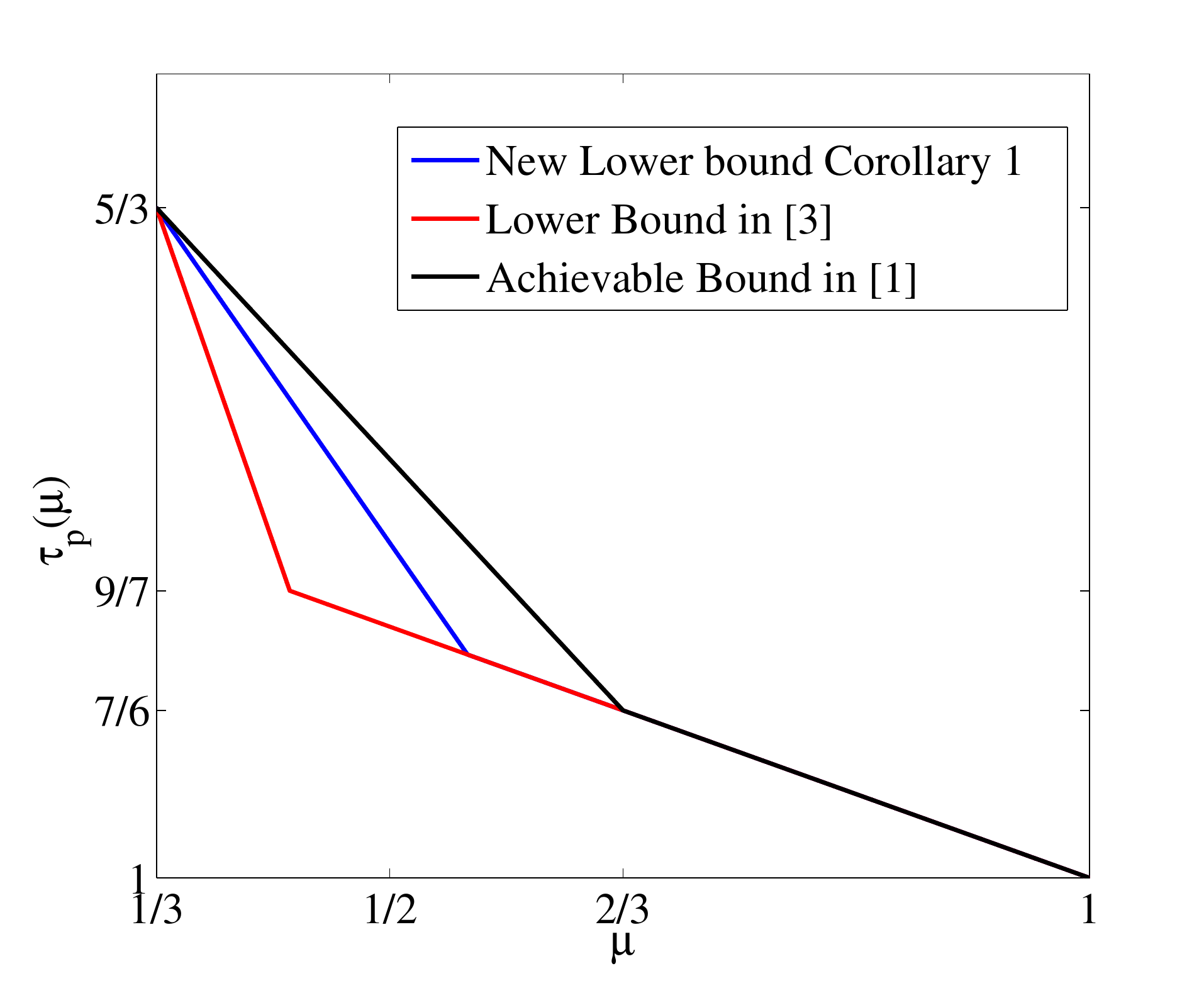}}
\caption{The peak NDT for a cache-aided interference network with $K_T=3$ transmitters and $K_R=3$ receivers.}
\label{fig2}
\vspace{-5mm}  
\end{figure}

 In the delivery phase, receiver $\text{RX}_j$ requests a file $W_{d_j}$ out of the $N$ files of the library. We consider $\mathbf{d}=\left[d_1,\cdots,d_{K_R}\right]\in\left[N\right]^{K_R}$ as the vector of receivers demands. The transmitters are informed with receivers demands $\mathbb{d}$. Thus, transmitter $\text{TX}_i$, $i\in\left[K_T\right]$, responds to the user demands by sending a codeword $\mathbf{x}_i\triangleq\left(x_i\left(t\right)\right)_{t=1}^{T}$ of block length $T$ over the interference channel, where $x_i\left(t\right)\in\mathbb{C}$ is the transmitted signal of transmitter $\text{TX}_i$ at time $t\in\left[T\right]$. We impose an average transmit power constraint over the channel input $\frac{1}{T}||\mathbf{x}_i||\leq P$. In this phase, each transmitter has only access to its own cache contents, therefore, the codeword $\mathbf{x}_i$ of transmitter $\text{TX}_i$ is determined by an encoding function in the receivers demands $\mathbf{d}$, the cache contents $Z_i$, and the channel coefficients between TXs and RXs. Afterwards, each receiver $\text{RX}_j$ implements a decoding function to estimate the requested file $\hat{W}_{d_j}$ from the received signal $\mathbf{y}_j\triangleq\left(y_j\left(t\right)\right)_{t=1}^{T}$ given by
\begin{equation}~\label{eqn1}
 y_j\left(t\right)=\sum_{i=1}^{K_T}h_{ji}x_i\left(t\right)+n\left(t\right)
\end{equation}
where $y_j\left(t\right)\in\mathbb{C}$ is the received signal by receiver $\text{RX}_j$ at time $t\in\left[T\right]$, and $n\left(t\right)$ denotes the additive white Gaussian noise at receiver $\text{RX}_j$. $h_{ji}\in\mathbb{C}$ represents the channel gain between transmitter $\text{TX}_i$ and receiver $\text{RX}_j$. Let $S\left(\mathbf{d}\right)$ be a function returning the number of distinct files in the demand $\mathbf{d}$. For a given demand $\mathbf{d}$, the system performance can be characterized by the normalized delivery time (NDT) defined as~\cite{tandon2016cloud}.
\begin{equation}
\tau\left(\mu,\mathbf{d}\right)=\lim_{P\to \infty}\lim_{F\to \infty}\frac{T\left(\mu,P,\mathbf{d}\right)}{F/\log\left(P\right)},
\end{equation}
where $T\left(\mu,P,\mathbf{d}\right)$ denotes the time needed to send the all requested files such that each receiver can decode its requested file with probability one as $F\to\infty$. The NDT refers to the delivery latency with respect to an interference-free baseline system at the high SNR regime. Furthermore, we define $\overline{\tau}\left(\mu\right)=\mathbb{E}_{\mathbf{d}}\left[\tau\left(\mu,\mathbf{d}\right)\right]$ as the expected NDT, where the expectation is over the random demand $\mathbf{d}$.

 Our objective in this work is to derive an information theoretic lower bound on the expected NDT as a function of the normalized cache size $\mu$ for cache-aided interference networks. We point out that the transmitter cache size must satisfy $K_T\mu\geq 1$ to maintain that every bit of the library content is stored at least at one cache of the network. Moreover, if the cache size increases the library size $\mu>1$, each transmitter is able to cache all the library files and the remaining cache memory would not be used. Therefore, we are interested in the normalized cache size $\frac{1}{K_T}\leq \mu\leq 1$.

\section{Main Results}

In this section, we first present our main result of this paper which gives a lower bound on the expected NDT for cache-aided networks. Then, for a special case when each receiver requests a distinct file, we compare our results with the cut-set based lower bound in~\cite[Theorem~1]{sengupta2016cache}.

\begin{theorem}~\label{Th1} For a $K_T\times K_R$ cache-aided interference network with a library of $N$ files, normalized cache size $\mu\in\left[\frac{1}{K_T}:1\right]$ at each transmitter, and a parameter $t=K_T\mu$, the expected NDT under uniform popularity distribution is lower bounded as
\begin{equation}
\overline{\tau}\left(\mu\right)\geq\mathbb{E}\left[\max_{\mathcal{F}}\text{Conv}\left( \frac{t{K_T\choose t}+\left(S\left(\mathbf{d}\right)-\sigma\right){\sigma-1\choose t-1}}{t{K_T\choose t}}\right)\right] 
\end{equation}
where $\mathcal{F}\triangleq \left\{1\leq\sigma\leq\min\lbrace K_T,S\left(\mathbf{d}\right)\rbrace\right\}$, and the expectation is over the random demand $\mathbf{d}$. $\text{Conv}\left(f\left(t\right)\right)$ denotes the lower convex envelope of the integer points $\left[\left(t,f\left(t\right)\right): t\in\lbrace 1,\cdots,K_T\rbrace \right]$.
\end{theorem}
To the best of our knowledge, this theorem gives the first converse bound on the \textit{expected} NDT under uniform popularity distribution for cache-aided interference networks, where the lower bound in~\cite[Theorem~1]{sengupta2016cache} is applied to the peak NDT only wherein each receiver requests a different file. To prove Theorem~\ref{Th1}, we first derive a lower bound on the NDT for a given demand $\mathbf{d}$, and uncoded placement scheme. The derived lower bound is mainly based on genie-aided, cut-set arguments. Then, we optimize the derived bound over all possible uncoded placement schemes to get the minimum NDT for a given demand $\mathbf{d}$. Finally, by taking the expectation over all demands $\mathbf{d}\in\left[N\right]^{K_R}$, we obtain the lower bound in Theorem~\ref{Th1}. The full proof of Theorem~\ref{Th1} is presented in Section~\ref{Sec4}. We can directly derive a lower bound on the peak NDT from Theorem~\ref{Th1} as in the following corollary.

\begin{corollary}~\label{Cor1} For a general $K_T\times K_R$ cache-aided interference network with a library of $N\geq K_R$ files, normalized cache size $\mu\in\left[\frac{1}{K_T}:1\right]$ at each transmitter, a parameter $t=K_T\mu$, and each receiver requests a distinct file, the NDT is lower bounded as \small
\begin{equation}
\tau_{\text{p}}\left(\mu\right)\geq\max_{1\leq\sigma\leq\min\left\{K_T,K_R\right\}}\text{Conv}\left(\frac{t{K_T\choose t}+\left(K_R-\sigma\right){\sigma-1\choose t-1}}{t{K_T\choose t}}\right)
\end{equation}

\normalsize 
\end{corollary}

\begin{figure}[t]
\centering
 \begin{subfigure}[b]{0.49\linewidth}
  \centerline{\includegraphics[scale=0.24,trim=4 4 4 4,clip]{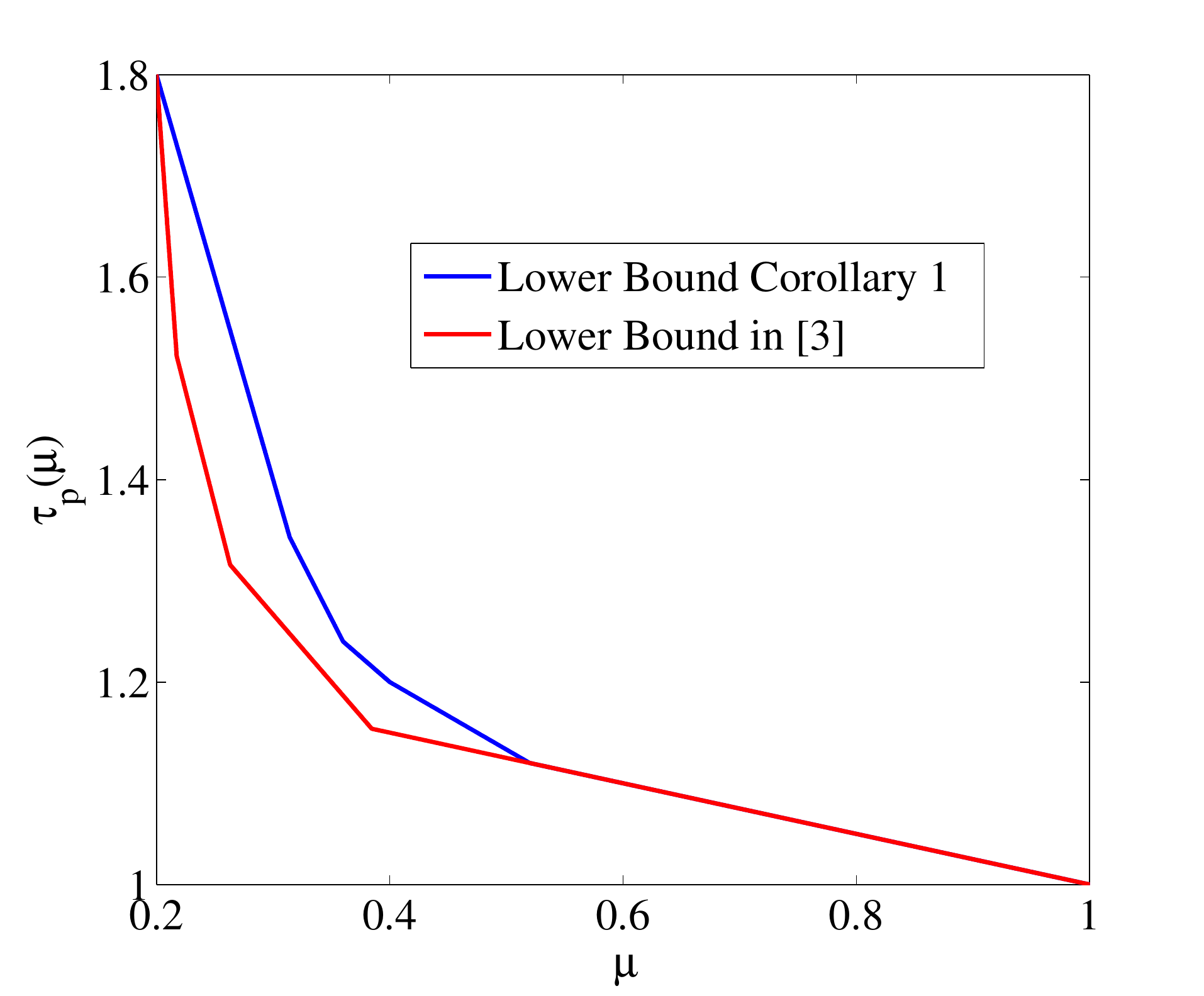}}
 \caption{$K_T=5$ and $K_R=5$.}\label{fig3A}
 \end{subfigure}
 \begin{subfigure}[b]{0.49\linewidth}
  \centerline{\includegraphics[scale=0.24,trim=4 4 4 4,clip]{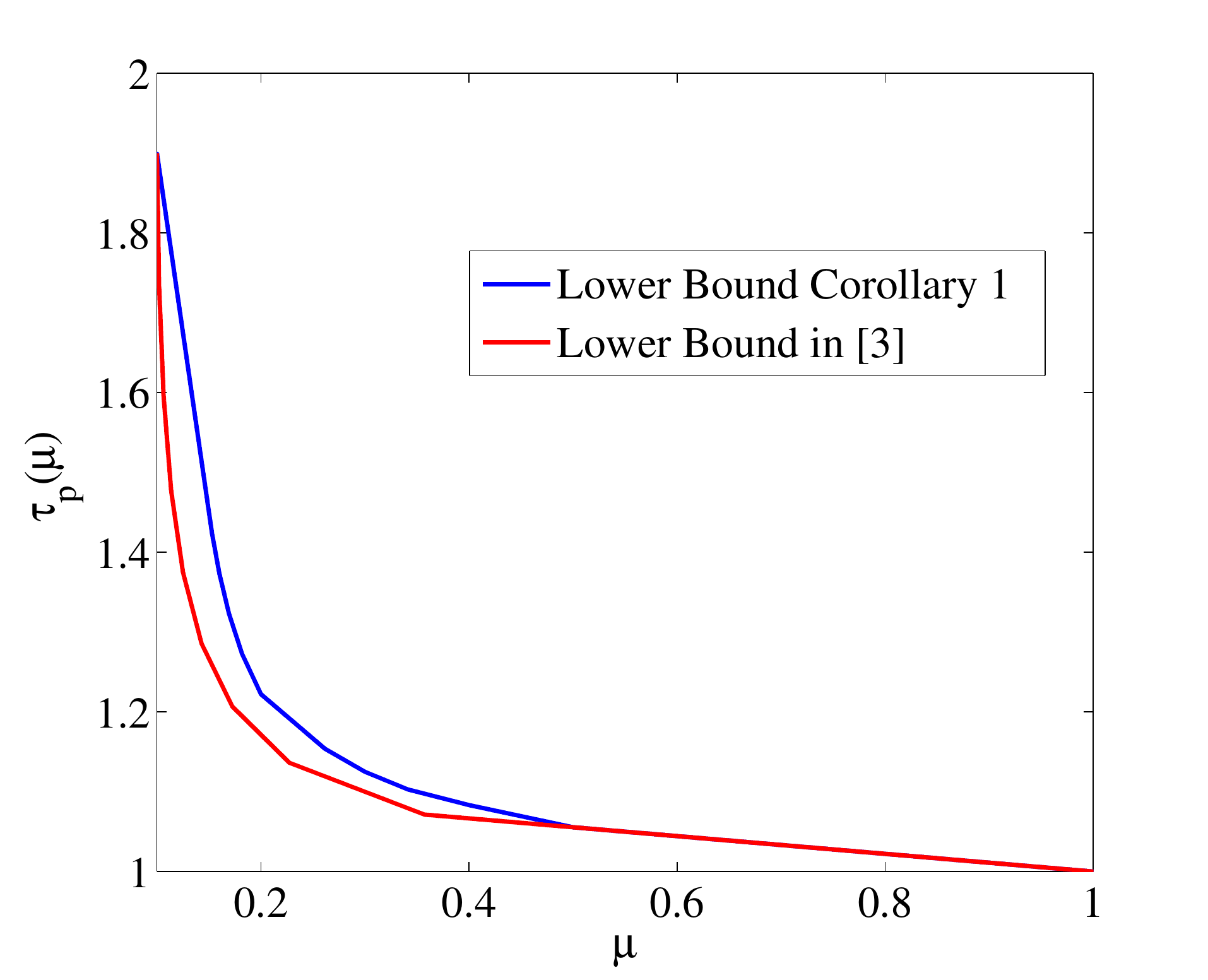}}
\caption{$K_T=10$ and $K_R=10$.}\label{fig3B}
 \end{subfigure}
\caption{Comparison between our bound in Corollary~\ref{Cor1} and the bound in~\cite{sengupta2016cache} for the peak NDT.}
\vspace{-5mm}
\label{Fig3}
\end{figure}

The proof is straightforward obtained from Theorem~\ref{Th1} by setting the number of distinct demands $S\left(\mathbf{d}\right)=K_R$. Now, we compare our result in Corollary~\ref{Cor1} with the lower bound in~\cite{sengupta2016cache}. In Figure~\ref{fig2}, we plot Maddah-Ali-Neisen (MN) scheme in~\cite{maddah2015cache}, the lower bound derived in~\cite{sengupta2016cache}, and our proposed lower bound in Corollary~\ref{Cor1} for a cache-aided interference network with $K_T=3$ transmitters and $K_R=3$ receivers. We can see that our bound is tighter than the bound in~\cite{sengupta2016cache} for $\mu\leq 0.5$, where the multiplicative gap between the MN scheme and our lower bound is reduced to $1.091$. In Figures~\ref{fig3A} and~\ref{fig3B}, we compare between our bound in Corollary~\ref{Cor1} and the bound in~\cite{sengupta2016cache} with different number of transmitters and receivers. It is shown that our bound is tighter when the normalized cache size $\mu\leq 0.5$, while our bound coincides with the bound in~\cite{sengupta2016cache} for large cache sizes when $\mu\geq 0.5$.

\begin{figure}[t]
\centering{\includegraphics[scale=0.26,trim=4 4 4 4,clip]{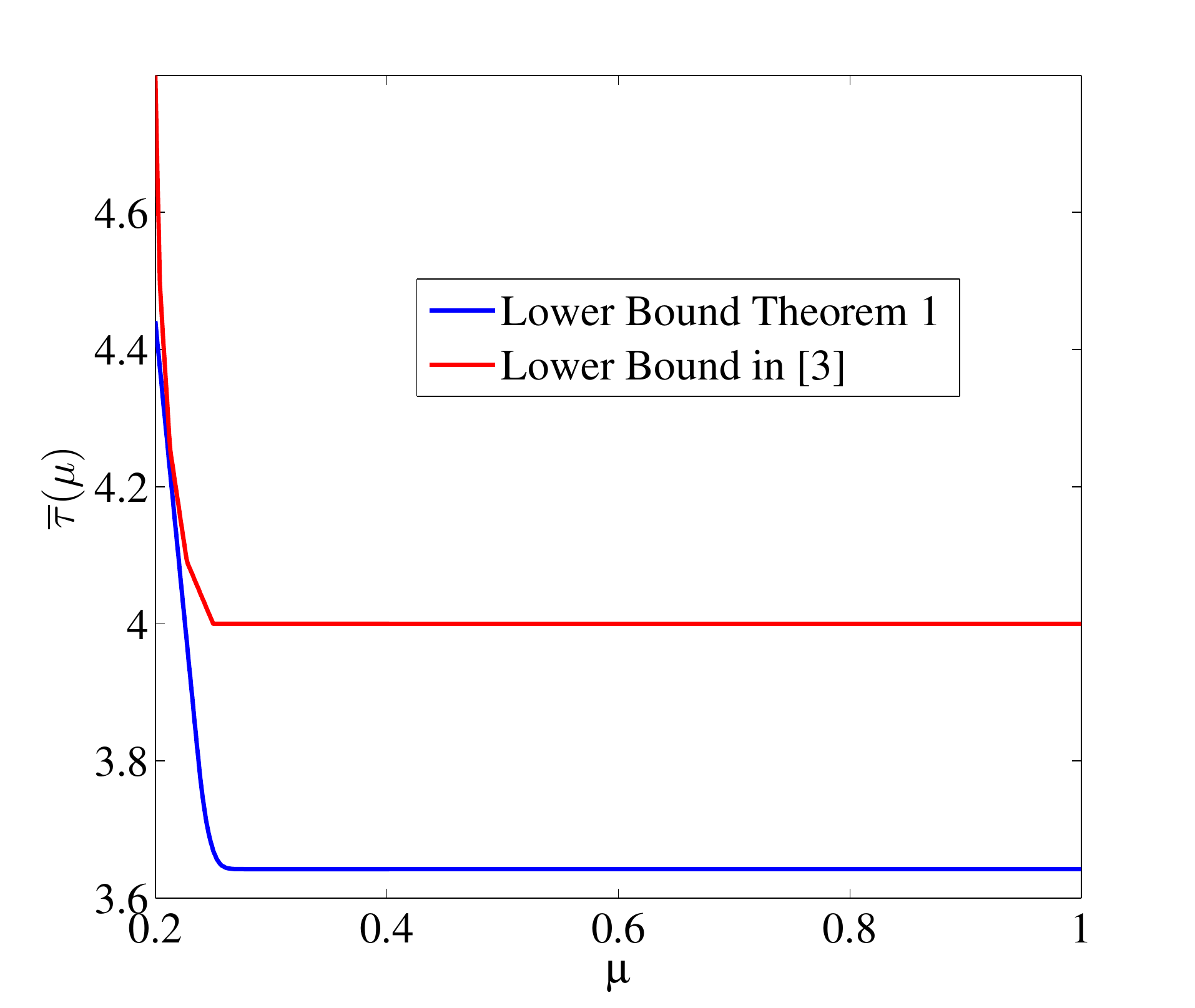}}
\caption{Converse bound on the expected NDT for a cache-aided interference network with $K_T=5$ transmitters, $K_R=20$ receivers, and a library of $N=100$.}
\label{fig4}
\vspace{-3mm}  
\end{figure}

In Figure~\ref{fig4}, we plot the lower bound on the expected NDT in Theorem~\ref{Th1} and the lower bound on the peak NDT in~\cite{sengupta2016cache} for a cache-aided interference network with $K_T=5$ transmitters, $K_R=20$ receivers, and a library of $N=100$ file. The expected NDT works differently from the peak NDT. In the peak NDT, each receiver requests a different file. Therefore, at each time, there will be $K_R$ different files required to be delivered, while in the expected NDT, there is a redundancy in the receivers requests, i.e., there is a chance that different receivers request the same file. Hence, it is expected that the NDT would be reduced. To see this consider a simple example of a single transmitter. For an extreme case when all receivers request the same file, the transmitter can broadcast this file in a single time slot to all receivers, i.e., $\tau=1$. While in the worst case, it is required $K_R$ time slots to send the different $K_R$ files, i.e., $\tau=K_R$. This interprets why our bound on the expected NDT is less than the bound on the peak NDT in~\cite{sengupta2016cache}. Moreover, this observation indicates that the feasibility region on the expected NDT is bigger than the feasibility region on the peak NDT, and hence, it is expected that the achievable schemes for the worst case demand might be no longer order optimal in general.


\section{Proof of Theorem~\ref{Th1}}~\label{Sec4}

In this section, we present the detailed proof of Theorem~\ref{Th1}. Let $\tau\left(\mathbf{d},\mu,\mathcal{Z}\right)$ denote the NDT for a given demand $\mathbf{d}$ and placement scheme $\mathcal{Z}\triangleq\left\{Z_1,\cdots,Z_{K_T}\right\}$. Then, the expected NDT can be bounded by
\begin{equation}
\begin{split}
\overline{\tau}\left(\mu\right)&=\min_{\mathcal{Z}}\mathbb{E}_{\mathbf{d}}\left[\tau\left(\mathbf{d},\mu,\mathcal{Z}\right)\right]\\
&\stackrel{\left(a\right)}{=}\min_{\mathcal{Z}}\mathbb{E}_{S\left(\mathbf{d}\right)}\left[\mathbb{E}_{\mathbf{d}|S\left(\mathbf{d}\right)}\left[\tau\left(S\left(\mathbf{d}\right),\mu,\mathcal{Z}\right)\right]\right]\\
&\stackrel{\left(b\right)}{\geq}\mathbb{E}_{S\left(\mathbf{d}\right)}\left[\min_{\mathcal{Z}}\mathbb{E}_{\mathbf{d}|S\left(\mathbf{d}\right)}\left[\tau\left(S\left(\mathbf{d}\right),\mu,\mathcal{Z}\right)\right]\right]
\end{split}
\end{equation}
where in step $\left(a\right)$, we first take the expectation over demands on condition that $S\left(\mathbf{d}\right)=s$, i.e., the number of distinct files in demand $\mathbf{d}$ is equal to $s$. Then, we take the expectation over all values of $s$. Notice that $\mathbf{d}$ is a random vector, and hence, $S\left(\mathbf{d}\right)$ is a random variable taking values from $\lbrace1,\cdots,\min\lbrace K_R,N\rbrace\rbrace$. Thus, we divide the demands $\mathbf{d}\in\left[N\right]^{K_R}$ into categories $\lbrace \mathcal{D}_s\rbrace$, where $\mathcal{D}_s$ is the set of demands satisfying that $S\left(\mathbf{d}\right)=s$, i.e., the demands that have exactly $s$ distinct files. In step $\left(b\right)$, we bound the expected NDT by designing the placement scheme to minimize individually the NDT for each demand category instead of designing the placement scheme to minimize the expected NDT. 

To obtain the result in Theorem~\ref{Th1}, we derive a lower bound on the NDT for demand category $\mathcal{D}_s$ by using cut-set and genie-aided arguments. Then, we run an optimization problem to find the tight cut over all possible cuts, and to minimize the NDT over all possible uncoded placement schemes. Finally, we take the expectation with respect to $S\left(\mathbf{d}\right)$.
%

For a given demand $\mathbf{d}\in\mathcal{D}_s$, let $\mathcal{R}$ be an arbitrary set of $S\left(\mathbf{d}\right)$ receivers, in which each receiver requests a different file. Let $\mathcal{S}_t$ be a set of transmitters with cardinality $\sigma$, and $\mathcal{S}_r$ be a set of receivers with cardinality $\sigma$, where $1\leq \sigma\leq s$. We define $\overline{\mathcal{S}}_t=\left[K_T\right]\setminus \mathcal{S}_t$, and $\overline{\mathcal{S}}_r=\mathcal{R}\setminus \mathcal{S}_r$. The cache contents of set $\mathcal{S}_t$ of transmitters is defined by $\mathcal{Z}_{\mathcal{S}_t}\triangleq\lbrace Z_i\rbrace_{i\in\mathcal{S}_t}$. Moreover, we define the following disjoint set of bits
\begin{equation}
\begin{aligned}
&\mathcal{W}_{\mathcal{S}_t}&&\triangleq\left\{B_{d_j,i}:B_{d_j,i}\notin\mathcal{Z}_{\overline{\mathcal{S}}_t}, j\in\mathcal{R}\right\}\\
&\mathcal{W}_{\mathcal{S}_r}&&\triangleq\left\{B_{d_j,i}:B_{d_j,i}\in\mathcal{Z}_{\overline{\mathcal{S}}_t}, j\in\mathcal{S}_r\right\}\\
&\overline{\mathcal{W}}&&\triangleq\left\{B_{d_j,i}:B_{d_j,i}\in\mathcal{Z}_{\overline{\mathcal{S}}_t}, j\in\overline{\mathcal{S}}_r\right\}
\end{aligned}
\end{equation}  
where $B_{d_j,i}$ denotes the $i$th bits in file $W_{d_j}$, for all $i\in\left[F\right]$. Observe that each bit of the library should be stored at least at one of the transmitter caches. Hence, if $B_{d_j,i}\notin\mathcal{Z}_{\overline{\mathcal{S}}_t}$, then $B_{d_j,i}\in\mathcal{Z}_{\mathcal{S}_t}$. The set $\mathcal{W}_{\mathcal{S}_t}$ contains the bits of files $\lbrace W_{d_j}\rbrace_{j\in\mathcal{R}}$ that are stored exclusively at the caches of transmitters $\mathcal{S}_t$, while the set $\mathcal{W}_{\mathcal{S}_r}$ contains the bits of files $\lbrace W_{d_j}\rbrace_{j\in\mathcal{S}_r}$ that are available at transmitters $\overline{\mathcal{S}}_t$. We can easily verify that $\mathcal{W}_{\mathcal{S}_t}\bigcup \mathcal{W}_{\mathcal{S}_r}$ has all the bits of files  $\lbrace W_{d_j}\rbrace_{j\in\mathcal{S}_r}$ in addition to the bits of files  $\lbrace W_{d_j}\rbrace_{j\in\overline{\mathcal{S}}_r}$ that are exclusively stored at transmitters $\mathcal{S}_t$. 

Assume that a genie provides the receivers in set $\mathcal{S}_r$ with the bits in set $\overline{\mathcal{W}}$, and provides the receivers in set $\overline{\mathcal{S}}_r$ with bits in set $\mathcal{W}_{\mathcal{S}_r}\bigcup\overline{\mathcal{W}}$. We prove that the set $\mathcal{S}_r$ of $\sigma$ receivers can decode all bits $\mathcal{W}_{\mathcal{S}_t}\bigcup \mathcal{W}_{\mathcal{S}_r}$ using their received signal and the genie-aided information. Consider the receivers in set $\mathcal{S}_r$ can fully cooperate between each others. We present the received signals of $\mathcal{S}_r$ and $\overline{\mathcal{S}}_r$ receivers as follows:
\begin{equation}
\begin{aligned}
&\mathbf{Y}_{\mathcal{S}_r}&=\mathbf{H}_{\mathcal{S}_r}^{\mathcal{S}_t}\mathbf{X}_{\mathcal{S}_t}+\mathbf{H}_{\mathcal{S}_r}^{\mathcal{\overline{S}}_t}\mathbf{X}_{\mathcal{\overline{S}}_t}+\mathbf{Z}_{\mathcal{S}_r},\\
&\mathbf{Y}_{\mathcal{\overline{S}}_r}&=\mathbf{H}_{\mathcal{\overline{S}}_r}^{\mathcal{S}_t}\mathbf{X}_{\mathcal{S}_t}+\mathbf{H}_{\mathcal{\overline{S}}_r}^{\mathcal{\overline{S}}_t}\mathbf{X}_{\mathcal{\overline{S}}_t}+\mathbf{Z}_{\mathcal{\overline{S}}_r}.\\
\end{aligned}
\end{equation}
where $\mathbf{Y}_{\mathcal{K}_r}$ is a $\left|\mathcal{K}_r\right|\times 1$ concatenated vector of the received signals of receivers in set $\mathcal{K}_r$, and $\mathbf{X}_{\mathcal{K}_t}$ is a $\left|\mathcal{K}_t\right|\times 1$ concatenated vector of the transmitted signals of transmitters in set $\mathcal{K}_t$. Furthermore, $\mathbf{H}_{\mathcal{K}_r}^{\mathcal{K}_t}=\left[h_{ji}\right]_{j\in\mathcal{K}_r}^{i\in\mathcal{K}_t}$ is a $\left|\mathcal{K}_r\right|\times\left|\mathcal{K}_t\right|$ channel matrix between transmitters in set $\mathcal{K}_t$ and receivers in set $\mathcal{K}_r$. For any coding scheme, receivers in $\mathcal{S}_r$ should be able to decode the bits $\mathcal{W}_{\mathcal{S}_r}$. Therefore, receivers in $\mathcal{S}_r$ can compute $\mathbf{X}_{\overline{\mathcal{S}}_t}=\lbrace x_i\rbrace_{i\in\overline{\mathcal{S}}_t}$ and subtract it from the received signal using the decoded bits $\mathcal{W}_{\mathcal{S}_r}$ and the genie-aided information $\mathcal{\overline{W}}$, where the encoding function of the transmitters are as follows
\begin{equation}
x_i=f_i\left(B_{d_j,l}:j\in\mathcal{R}, B_{d_j,l}\in Z_i\right). 
\end{equation}
Similarly, receivers in set $\overline{\mathcal{S}}_r$ can compute $\mathbf{X}_{\overline{\mathcal{S}}_t}$ and subtract it from the received signal using the genie-aided information $\mathcal{W}_{\mathcal{S}_r}\bigcup\overline{\mathcal{W}}$. As a result, we can rewrite the received signals of receivers in $\mathcal{S}_r$ and $\overline{\mathcal{S}}_r$ as
\begin{equation}
\begin{aligned}
&\mathbf{\tilde{Y}}_{\mathcal{S}_r}&=\mathbf{H}_{\mathcal{S}_r}^{\mathcal{S}_t}\mathbf{X}_{\mathcal{S}_t}+\mathbf{Z}_{\mathcal{S}_r},\\
&\mathbf{\tilde{Y}}_{\mathcal{\overline{S}}_r}&=\mathbf{H}_{\mathcal{\overline{S}}_r}^{\mathcal{S}_t}\mathbf{X}_{\mathcal{S}_t}+\mathbf{Z}_{\mathcal{\overline{S}}_r}.\\
\end{aligned}
\end{equation}
where receivers $j\in\mathcal{\overline{S}}_r$ are able to decode their bits $\lbrace B_{d_j,i}:B_{d_j,i}\notin\mathcal{Z}_{\overline{\mathcal{S}}_t},j\in\mathcal{\overline{S}}_r\rbrace$ from the received signal vector $\mathbf{\tilde{Y}}_{\mathcal{\overline{S}}_r}$, and receivers $j\in\mathcal{S}_r$ are able to decode their intended bits $\lbrace B_{d_j,i}:B_{d_j,i}\notin\mathcal{Z}_{\overline{\mathcal{S}}_t},j\in\mathcal{S}_r\rbrace$ from the received signal vector $\mathbf{\tilde{Y}}_{\mathcal{S}_r}$. Notice that the $\sigma \times \sigma $ submatrix channel $\mathbf{H}_{\mathcal{S}_r}^{\mathcal{S}_t}$ is invertable almost surely. Thus, by reducing noise at receivers $\mathcal{S}_r$ and multiplying the constructed signal $\mathbf{\tilde{Y}}_{\mathcal{S}_r}$ at receivers $\mathcal{S}_r$ by $\mathbf{H}_{\mathcal{\overline{S}}_r}^{\mathcal{S}_t}\left(\mathbf{H}_{\mathcal{S}_r}^{\mathcal{S}_t}\right)^{-1}$, we have 
\begin{equation}
\mathbf{\tilde{Y}}'_{\mathcal{S}_r}=\mathbf{H}_{\mathcal{\overline{S}}_r}^{\mathcal{S}_t}\mathbf{X}_{\mathcal{S}_t}+\mathbf{\tilde{Z}}_{\mathcal{S}_r},
\end{equation}  
which is  a degraded version of $\mathbf{\tilde{Y}}_{\mathcal{\overline{S}}_r}$, where $\mathbf{\tilde{Z}}_{\mathcal{S}_r}$ represents the reduced noise vector at receivers $\mathcal{S}_r$. Therefore, receivers in set $\mathcal{S}_r$ can decode all messages $\mathcal{W}_{\mathcal{S}_t}$. Thus, by using Fano's inequality, we have
\begin{equation}
H\left(\mathcal{W}_{\mathcal{S}_t}|\mathbf{Y}_{\mathcal{S}_r},\mathcal{\overline{W}}\right)\leq H\left(\mathcal{W}_{\mathcal{S}_t}|\mathbf{Y}_{\mathcal{\overline{S}}_r},\mathcal{\overline{W}},\mathcal{W}_{\mathcal{S}_r}\right)\leq |\mathcal{W}_{\mathcal{S}_t}|T\epsilon.
\end{equation}
The applied assumptions (genie-aided information, cooperation between subset of receivers, reducing noise) cannot hurt the coding scheme. Thus, we have
\begin{IEEEeqnarray}{lll}~\label{Lower1}
H&\left(\mathcal{W}_{\mathcal{S}_t},\mathcal{W}_{\mathcal{S}_r}\right)\\
=&\sum_{j\in\mathcal{S}_r}\sum_{i=1}^{F}\mathbf{1}\left(B_{d_j,i}\in\mathcal{Z}\right)+\sum_{j\in\mathcal{\overline{S}}_r}\sum_{i=1}^{F}\mathbf{1}\left(B_{d_j,i}\notin\mathcal{Z}_{\mathcal{\overline{S}}_t}\right) \nonumber\\
\stackrel{\left(a\right)}{=}&H\left(\mathcal{W}_{\mathcal{S}_t},\mathcal{W}_{\mathcal{S}_r}|\mathcal{\overline{W}}\right)\nonumber\\
\stackrel{\left(b\right)}{=}& I\left(\mathcal{W}_{\mathcal{S}_t},\mathcal{W}_{\mathcal{S}_r};\mathbf{Y}_{\mathcal{S}_r}|\mathcal{\overline{W}}\right)+H\left(\mathcal{W}_{\mathcal{S}_t},\mathcal{W}_{\mathcal{S}_r}|\mathbf{Y}_{\mathcal{S}_r},\mathcal{\overline{W}}\right)\\
\stackrel{\left(c\right)}{\leq}& I\left(\mathbf{X}_{\left[K_T\right]};\mathbf{Y}_{\mathcal{S}_r}\right)+H\left(\mathcal{W}_{\mathcal{S}_t},\mathcal{W}_{\mathcal{S}_r}|\mathbf{Y}_{\mathcal{S}_r},\mathcal{\overline{W}}\right)\nonumber\\
\stackrel{\left(d\right)}{\leq}& T\sigma\log\left(P\right)+H\left(\mathcal{W}_{\mathcal{S}_r}|\mathbf{Y}_{\mathcal{S}_r},\mathcal{\overline{W}}\right)+H\left(\mathcal{W}_{\mathcal{S}_t}|\mathbf{Y}_{\mathcal{S}_r},\mathcal{\overline{W}},\mathcal{W}_{\mathcal{S}_r}\right)\nonumber\\
\stackrel{\left(e\right)}{\leq}& T\sigma\log\left(P\right)+|\mathcal{S}_r|T\epsilon+|\mathcal{S}_t|T\epsilon\nonumber
\end{IEEEeqnarray} 
where $\mathbf{1}\left(.\right)$ is an indicator function. $\left(a\right)$ follows from the fact that the sets of bits are independent. Step $\left(b\right)$ follows from the chain rule. Step $\left(c\right)$ follows from data processing inequality, where the signal $\mathbf{X}_{\left[K_T\right]}$ is a function of $\mathcal{W}_{\mathcal{S}_t}\bigcup\mathcal{W}_{\mathcal{S}_r}$. Step $\left(d\right)$ follows from the bound of the degrees of freedom of multiple access channel (MAC) with $K_T$ single-antenna transmitters and a receiver with $|\mathcal{S}_r|$ antennas. Finally, step $\left(e\right)$ follows from Fano's inequality. By diving on $F$, and taking $P\to\infty$ and $\epsilon\to 0$, we get.
\begin{equation}
\begin{aligned}
\frac{1}{F}\left(\sum_{j\in\mathcal{S}_r}\sum_{i=1}^{F}\mathbf{1}\left(B_{d_j,i}\in\mathcal{Z}\right)+\sum_{j\in\mathcal{\overline{S}}_r}\sum_{i=1}^{F}\mathbf{1}\left(B_{d_j,i}\notin\mathcal{Z}_{\mathcal{\overline{S}}_t}\right)\right)\\ 
\leq \sigma\tau\left(\mu,\mathbf{d},\mathcal{Z}\right).
\end{aligned}
\end{equation}
Notice $\mathbf{1}\left(B_{d_j,i}\in\mathcal{Z}\right)=1$ for any bit in the library, since every bit should be available at least at one of the transmitter caches. Hence, the first term in the left hand side (LHS) is equal to $\sigma F$. Then, by taking the average of the above inequality over all possible set $\mathcal{S}_r\subset\mathcal{R}$, we have
\small
\begin{equation}
\sigma+\frac{{s-1\choose s-\sigma-1}}{F{s\choose \sigma}}\left(\sum_{j\in\mathcal{R}}\sum_{i=1}^{F}\mathbf{1}\left(B_{d_j,i}\notin\mathcal{Z}_{\mathcal{\overline{S}}_t}\right)\right)\leq\sigma \tau\left(\mu,\mathbf{d},\mathcal{Z}\right).
\end{equation}
\normalsize where every indicator $\mathbf{1}\left(B_{d_j,i}\notin\mathcal{Z}_{\mathcal{\overline{S}}_t}\right)$ in the second term in the LHS is counted ${s-1\choose s-\sigma-1}$ times. Now, we follow similar steps as in~\cite{yu2017exact} to average the above inequality over all possible demands $d\in\mathcal{D}_s$, and all possible transmitter sets. Let $\mathcal{K}_{d_j,i}$ denote the set of transmitters that exclusively store the $i$-th bit of the file $W_{d_j}$. Thus, $\mathbf{1}\left(B_{d_j,i}\notin\mathcal{Z}_{\mathcal{\overline{S}}_t}\right)=\mathbf{1}\left(\mathcal{K}_{d_j,i}\bigcap\mathcal{\overline{S}}_t=\phi\right)$. By taking the average of all possible set $\mathcal{S}_t\subset\left[K_T\right]$, the second term in the LHF is equal
\begin{equation}~\label{eqn4}
\frac{s-\sigma}{Fs}\left(\sum_{j\in\mathcal{R}}\sum_{i=1}^{F}\frac{\sum\limits_{\mathcal{S}_t\subset\left[K_T\right]}\mathbf{1}\left(\mathcal{K}_{d_j,i}\bigcap\mathcal{\overline{S}}_t=\phi\right)}{{K_T\choose \sigma}}\right).
\end{equation}
where \small$\frac{{s-1\choose s-\sigma-1}}{{s\choose \sigma}}=\frac{s-\sigma}{s}$\normalsize, and we exchange the order of summations. The term \small$\frac{1}{{K_T\choose \sigma}}\sum\limits_{\mathcal{S}_t\subset\left[K_T\right]}\mathbf{1}\left(\mathcal{K}_{d_j,i}\bigcap\mathcal{\overline{S}}_t=\phi\right)$ \normalsize is equal to the probability of selecting $K_T-\sigma$ transmitters uniformally at random, and none of them belongs to $\mathcal{K}_{d_j,i}$. Hence, this term can be computed as follows\footnote{We assume that ${n\choose k}=0$ if $n<k$.}
\begin{equation}~\label{eqn2}
\frac{1}{{K_T\choose \sigma}}\sum\limits_{\mathcal{S}_t\subset\left[K_T\right]}\mathbf{1}\left(\mathcal{K}_{d_j,i}\bigcap\mathcal{\overline{S}}_t=\phi\right)=\frac{{{K_T-|\mathcal{K}_{d_j,i}|}\choose {K_T-\sigma}}}{{K_T\choose {K_T-\sigma}}}.
\end{equation} 
Let $a_{n,d_j}$ denote the number of bits of file $W_{d_j}$ that are stored exclusively at $n$ transmitters, and hence, $|\mathcal{K}_{d_j,i}|=n$ for a fraction $a_{{n,d_j}}/F$. By taking the average~\eqref{eqn2} over all bits of file $W_{d_j}$, we obtain
\begin{equation}~\label{eqn3}
\frac{1}{F}\sum_{i=1}^{F}\frac{{{K_T-|\mathcal{K}_{d_j,i}|}\choose {K_T-\sigma}}}{{K_T\choose {K_T-\sigma}}}=\sum_{n=1}^{K_T}\frac{a_{n,d_j}}{F}\frac{{{K_T-n}\choose {K_T-\sigma}}}{{K_T\choose {K_T-\sigma}}}=\sum_{n=1}^{K_T}\frac{a_{n,d_j}}{F}\frac{{\sigma\choose n}}{{K_T\choose n}}
\end{equation}
where we use the equality ${{K-n}\choose l}/{K\choose l}={{K-l}\choose n}/{K\choose n}$. Substituting from~\eqref{eqn3} into~\eqref{eqn4}, then we obtain
\begin{equation}
1+\frac{s-\sigma}{\sigma s}\sum_{j\in\mathcal{R}}\sum_{n=1}^{K_T}\frac{a_{n,d_j}}{F}\frac{{\sigma\choose n}}{{K_T\choose n}}\leq\tau\left(\mu,\mathbf{d},\mathcal{Z}\right).
\end{equation}
By taking the average over demands $\mathbf{d}\in\mathcal{D}_s$, we get
\begin{equation}~\label{eqn5}
1+\frac{s-\sigma}{\sigma s}\frac{1}{|\mathcal{D}_s|}\sum_{\mathbf{d}\in\mathcal{D}_s}\sum_{j\in\mathcal{R}}\sum_{n=1}^{K_T}\frac{a_{n,d_j}}{F}\frac{{\sigma\choose n}}{{K_T\choose n}}\leq\overline{\tau}\left(\mu,s\left(\mathbf{d}\right),\mathcal{Z}\right).
\end{equation}
It is easy to verify that demands $\mathbf{d}\in\mathcal{D}_s$ are uniformally distributed, since $\mathbf{d}\in\left[N\right]^{K_R}$ is a random vector with uniform distribution. Moreover, for file $W_j$, $j\in\left[N\right]$, the term $a_{n,j}$ is computed ${{N-1}\choose {s-1}}|\mathcal{D}_s|/{{N}\choose {s}}$ times in the summation $\sum_{\mathbf{d}\in\mathcal{D}_s}\sum_{j\in\mathcal{R}}\frac{a_{n,d_j}}{F}$. Thus,~\eqref{eqn5} is equal to
\begin{equation}~\label{eqn7}
1+\frac{s-\sigma}{\sigma}\sum_{j=1}^{N}\sum_{n=1}^{K_T}\frac{a_{n,j}}{NF}\frac{{\sigma\choose n}}{{K_T\choose n}}\leq\overline{\tau}\left(\mu,s\left(\mathbf{d}\right),\mathcal{Z}\right).
\end{equation}
Let $\alpha_n=\sum_{j=1}^{N}a_{n,j}/NF$, and $K_T\mu=t$. By minimizing both sides of~\eqref{eqn7} over all possible uncoded placement schemes, we get
\begin{equation}~\label{eqn6}
\begin{split}
1+&\frac{s-\sigma}{\sigma}\min_{\mathcal{Z}}\sum_{n=1}^{K_T}\alpha_n\frac{{\sigma\choose n}}{{K_T\choose n}}\leq\min_{\mathcal{Z}}\overline{\tau}\left(\mu,s\left(\mathbf{d}\right),\mathcal{Z}\right)\\
s.t.&\qquad \sum_{n=1}^{K_T}\alpha_{n}=1\\
 &\qquad \sum_{n=1}^{K_T}n\alpha_{n}=t
\end{split}
\end{equation} 
where the first constraint comes from the total  number of bits in library, while the second constraint is to maintain the total size of transmitter caches. Notice that $f_n=\frac{{\sigma\choose n}}{{K_T\choose n}}$ is a decreasing function of $n$. Moreover, we can verify that $f_n$ is a discrete convex function of $n$, since $f_{n+1}+f_{n-1}\geq 2f_{n}$ in region $1\leq n\leq\sigma$~\cite[Theorem~1]{murota1998discrete}. The objective function is a linear combination of points $\lbrace f_n\rbrace$. Hence, the optimal solution is $\alpha_t=1$ when $t$ is integer, i.e., $t\in\left[1:K_T\right]$. While for non-integer point of $t$, we can write $t=\alpha t_1+\left(1-\alpha\right) t_2$, where $t_1\leq t\leq t_2$. Thus, the optimal solution is $\alpha_{t_1}=\alpha$ and $\alpha_{t_2}=\left(1-\alpha\right)$. Therefore, we can proceed the proof to bound the expected NDT for the corner points $t\in\left[1:K_T\right]$, where the expected NDT for non-integer $t$ can be bounded  by the linear combination of the nearest two integer points. Thus, we get
\begin{equation}~\label{eqn8}
1+\frac{s-\sigma}{\sigma}\frac{{\sigma\choose t}}{{K_T\choose t}}\leq\min_{\mathcal{Z}}\overline{\tau}\left(\mu,s\left(\mathbf{d}\right),\mathcal{Z}\right)
\end{equation}
To get the best tight bound on the NDT, we maximize the LHS of~\eqref{eqn8} over all possible values of $\sigma\in\mathcal{F}\triangleq\lbrace 1\leq \sigma\leq \min\lbrace K_T,s\left(d\right)\rbrace\rbrace$.
\begin{equation}
\max_{\sigma\in\mathcal{F}}\frac{t{K_T\choose t}+\left(s-\sigma\right){\sigma-1\choose t-1}}{t{K_T\choose t}}\leq\min_{\mathcal{Z}}\overline{\tau}\left(\mu,s\left(\mathbf{d}\right),\mathcal{Z}\right).
\end{equation}
Finally, by taking the expectation with respect to $s\left(\mathbf{d}\right)$, we have
\begin{equation}
\mathbb{E}\left[\max_{\sigma\in\mathcal{F}}\frac{t{K_T\choose t}+\left(s-\sigma\right){\sigma-1\choose t-1}}{t{K_T\choose t}}\right]\leq\overline{\tau}\left(\mu\right).
\end{equation}
This completes the proof of Theorem~\ref{Th1}.

\section{Conclusion}
We have derived a lower bound on the \textit{expected} normalized delivery time for cache-aided interference networks under uniform popularity distribution. Our bound is mainly based on cut-set and genie-aided arguments. For peak NDT, the results have shown that our lower bound is tighter than the bound in~\cite{sengupta2016cache} for small cache sizes, while both bounds coincide with each other for large cache sizes. Furthermore, We have shown that the feasible region of the expected NDT is bigger than the feasible region of the peak NDT. Hence, the achievable schemes on the peak NDT might no longer becomes order optimal with respect to the new derived bound on the expected NDT.

\end{document}